\documentclass[prb,twocolumn,citeautoscript,groupedaddress]{revtex4}
\usepackage{graphicx}
\newcommand{\ud}{\,{\mathrm d}}

\newcommand{\citen}{\onlinecite}

\begin{document}
\title{Map of metastable states for thin circular magnetic nano-cylinders}
\author{Konstantin L. Metlov}
\affiliation{Donetsk Institute for Physics and Technology NASU, Donetsk 83114, Ukraine}
\email{metlov@kinetic.ac.donetsk.ua}
\author{YoungPak Lee}
\affiliation{Quantum Photonic Science Research Center and Dept. of Physics, Hanyang University, Seoul 133-791, Korea}
\date{\today}
\begin{abstract}
Nano-magnetic systems of artificially shaped ferromagnetic islands, recently became a popular subject due to their current and potential applications in spintronics\cite{Cowburn_NatM_2007}, magneto-photonics\cite{GAP05_magneto_photonics} and superconductivity\cite{GPMM04_dotqubit}. When the island size is close to the exchange length of magnetic material (around 15 nm), its magnetic structure becomes markedly different\cite{UP93,SOHSO00}. It determines both static and dynamic magnetic properties of elements, but strongly depends on their shape and size\cite{MG02_JEMS,SGNSSCF2003}. Here we map this dependence for circular cylindrical islands of a few exchange lengths in size. We outline the region of metastability of ``C''-type magnetic states, proving that they are indeed genuine and not a result of pinning on particle imperfections. A way to create the smallest particles with guaranteed magnetic vortex state at zero field becomes evident. It is expected that the map will help focus the efforts in planning of experiments and devices.
\end{abstract}
\maketitle
Since the first theoretical model for magnetic vortex in circular cylinder, taking into account long-range magnetic dipolar interaction\cite{UP93}, and experimental discovery of magnetic vortex core\cite{SOHSO00} there have been a number of developments in this area. The question of magnetic ground states in cylinders of various sizes was settled\cite{UP93,UP94,MG02_JEMS} and  thoroughly tested\cite{SGNSSCF2003} both experimentally and numerically. 

Yet, the most important property of ferromagnets, responsible for wealth of their present and future applications, is the ability to be in one of many metastable states. Fortunately, the number of distinct metastable states for nano-magnets is much smaller than in ordinary ones, making it feasible to enumerate and map them all. Such a map for circular cylinders in coordinates of $R/L_E$ vs $L/L_E$, where $L$ and $R$ are the particle's thickness and radius, respectively, and $L_E=\sqrt{4 \pi C/\mu_0 \gamma_B M_S^2}$ is the exchange length, is shown in Figure. $C$ and $M_S$ are the exchange constant and the saturation magnetization of material, respectively, $\mu_0$ is the permeability of vacuum in SI units or 1 in CGS and $\gamma_B$ is 1 in SI or $4\pi$ in CGS. Let us discuss the map first, before diving into the details on how the additional lines were derived.
\begin{figure*}
\label{fig:map}
 \includegraphics[scale=0.6]{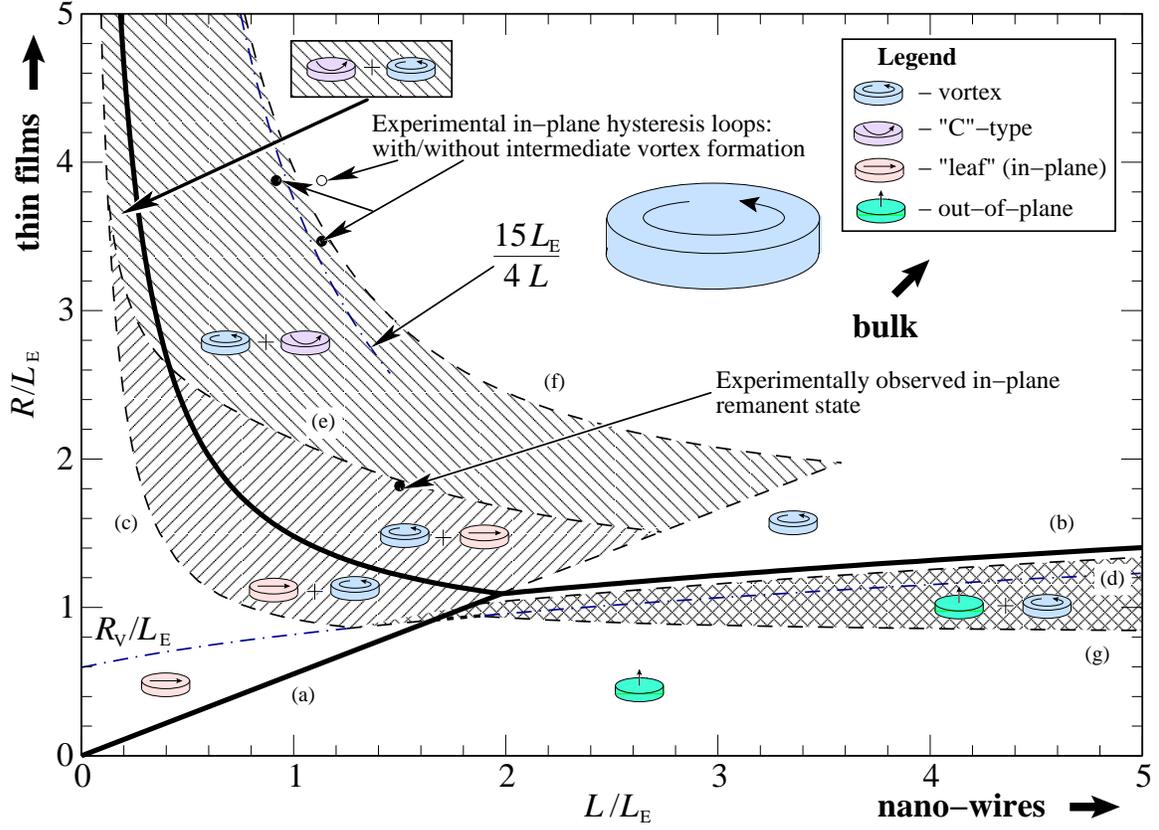}
\caption{Map of metastable circular nano-cylinders. Solid lines represent the equality of the equilibrium energies of the bordering ground states. Dashed lines correspond to the stability loss. More than one state (represented by symbols with arrows) can be stable in the shaded regions. The leftmost in each group of symbols indicates the ground state. The formula gives the asymptotic to the ``C''-vortex transition line, and the other dash-dot line is the equilibrium radius of the vortex core.}
\end{figure*}

The simplest magnetic states of nano-magnets are (quasi-)uniform ones: in-plane and out-of-plane. Neglecting quasi-uniformity, making only a small correction\cite{MG04}, the energies of these states\cite{J66} are equal in cylinders with aspect ratio $g = L/R = g_J \approx 1.8129523$. At this critical aspect ratio (straight line ``a'' in Figure) there is no energy barrier between the states, and the equilibrium line coincides with the stability line for each.

Solving for geometries with the energy of the vortex state equal to the energies of the (quasi-)uniform states (in-plane or out-of-plane, depending on aspect ratio), yields\cite{UP93,UP94} line ``b''.  Lines ``a'' and ``b'' separate the regions with different ground states of the particle, which can be reached experimentally, \textit{e.g.}, by annealing or thermal relaxation simply through waiting for a sufficient time.

Study of the vortex stability with respect to the displacement from the center of circular face produced\cite{MG02_JEMS} line ``c'' on the map. Below it (at a low $R/L_E$) the vortex state is impossible without external stabilization. This is the lower bound for the region of coexistence of the vortex and the (quasi-) uniform states. An upper bound of another metastability region, where the vortex state coexists with the out-of-plane uniform one\cite{GN04_JAP_PERP} is shown as line ``d,'' above which the uniform state is unstable. From general considerations the line ``b'' should coincide or lie below the line ``d''. It is not so (by hardly experimentally detectable amount) due to neglected magnetization variation along the cylinder thickness, as discussed in Ref.~\onlinecite{GN04_JAP_PERP}.

Here we fully outline the metastability regions by investigating the bending instability of the quasi-uniform ``leaf'' state (line ``e''), as well as stability of vortex with respect to the core expansion (line ``g''). It was puzzling at first to obtain the stability line ``e,'' crossing the equilibrium line ``b,'' but it turned out that destabilization of the ``leaf'' state does not imply the vortex formation. Another energy minimum appears, capturing the magnetization in a half-bent ``C'' state. Studying its stability yields line ``f'', where the capturing minimum vanishes and only the vortex state remains stable.

In-plane magnetization of ``C'' and ``leaf'' is unstable in elements with  $g<g_J$, giving rise to a ``dip'' of the vortex-only region down to the tricritical point of the diagram. If the in-plane state is created there (\textit{e.g.}, by a transient application of magnetic field), it turns into an out-of-plane state (because $g<g_J$) and collapses into vortex (since the region is above the line ``d'').

A number of experiments and numerical results\cite{SGNSSCF2003,RHSCea2002,CKAWT99} confirm this map. Let us present only few, supporting the new lines most directly. First, a ``unfitting''\cite{SGNSSCF2003} point from an experiment\cite{RHSCea2002}, where a remanent in-plane magnetization state was found in dots well above the equilibrium line ``b.'' It fits precisely in the metastability region for the ``leaf'' state. Even more direct confirmation comes from in-plane hysteresis measurements\cite{CKAWT99}, where in-plane hysteresis loops of supermalloy dots of different sizes were classified. The data show sketchily the boundary, separating the smaller dots with narrow rotational hysteresis, and bigger dots, where the loop is of nucleation type. It is exactly the stability line ``f''. Below the line, reduction of the field from in-plane saturation creates the ``C'' magnetization state, which is then rotated (as field passes zero) and straightened to become the ``leaf'' again. Above the line, the ``C'' state is unstable and vortex is nucleated instead, which must be annihilated, producing nucleation-type hysteresis loops. All the boundary points\cite{CKAWT99} fit the map, but only three of them, most tightly converged around the line ``f'' are plotted.

Let us now briefly describe the steps to obtain the lines. The starting point is provided by a recently developed approach\cite{M01_solitons2} to systematic generation of Ritz functions for magnetization distributions in thin nano-cylinders and its generalization\cite{MG04,M06}. We will not follow it here in detail, but only present the resulting function as an anzats
\begin{equation}
 \label{eq:fz}
 f(z)=\frac{\imath c z}{p}+\frac{1}{2}\left(a -\frac{z^2\overline{a}}{p^2}\right),
\end{equation} 
where $z=X/R+\imath Y/R$ is the normalized complex coordinate on the face of the circular cylinder. $c$ is real, $|z|<1$,  and $a$, $p$ are (in general) complex Ritz parameters, whose values are determined by minimizing the total energy of the particle (consisting of the exchange and the magnetostatic contributions). Normalized magnetization vector $\vec{m}=\vec{M}/M_S$ ($|\vec{m}|=1$) components are expressed as $m_X+\imath m_Y = 2 w /(1 + w \overline{w})$, $m_Z= (1 - w \overline{w})/(1 + w \overline{w})$, where a complex (but not holomorphic) function $w(z,\overline{z})$ is equal to $f(z)$ if $|f(z)|<1$ and $f(z)/|f(z)|$ if $f(z)\geq 1$.

All the variety of magnetization distributions in nano-scale circular dots is given by (\ref{eq:fz}) with different values of $c$, $p$ and $a$. When $p=1$, the magnetization has no normal components to the particle's circular side. If $|a|<c$, it is a displaced magnetic vortex. When $a=0$, the vortex is centered (the case studied by Usov and Peschany\cite{UP93} to find the vortex core radius $R_V = R/c$). Consideration of $0\leq|a| \ll 1$ yields\cite{MG02_JEMS} line ``c'' in Figure. When $c=0$ and $a=0$, the magnetization is uniform out-of-plane ($m_Z=1$, and $m_X=m_Y=0$). Expanding its energy at $0\leq c \ll 1$ allows to find the cylinder radius at which the out-of-plane configuration loses stability and is transformed into vortex\cite{GN04_JAP_PERP}. The case of $|p| > 1$, $c=0$ and $|a|>0$ corresponds to the quasi-uniform ``leaf'' configuration\cite{MG04}. When $p$ is imaginary, the configuration is changed into ``flower,'' which is not realized with isolated dots, but can be stabilized by interaction in dot arrays\cite{M06}.

In finding the lines ``e'' and ``f'' we permit non-zero values of $c$ in the quasi-uniform state ($p>1$ and $|a|>c$), bending the ``leaf''. The equation (\ref{eq:fz}) is parametrized by letting $c=x a$, where $|x|<1$, assuming that $p \gg 1$, $a$ is real and $a>a_0$, such that $|f(z)| > 1$ always. The magnetization is independent of $a$ and at $x=0$ coincides with that in Ref.~\citen{MG04} (noting that $x$ of Ref.~\citen{MG04} is called $p$ here). Evaluation of magnetostatic energy of this distribution is cumbersome (we use the magnetic charges formalism and factorization\cite{MG04}). Normalized to $\mu_0\gamma_B M_s^2 L \pi R^2$ up to the fourth order in $1/p$, it is
\begin{eqnarray}
 \nonumber
 e_{MS}^1\!=\!e_\parallel(g)\!+\!\frac{e_2(g)\!+\!x^2 e_{22}(g)}{p^2}\!+\!\frac{e_4(g)\!+\!x^2 e_{42}(g)\!+\!x^4 e_{44}(g)}{p^4}.
\end{eqnarray}
The functions $e^\parallel$, $e_2$, and $e_4$ are exactly the same as in Ref.~\citen{MG04} (to which expression this new result passes in a limit), and the three new functions are
\begin{eqnarray}
 \nonumber
  e_{22}(g)\!\!&=&\!\!\!\!\int\limits_0^\infty\!\frac{F(k g)}{k}\frac{J_2^2(k)-J_1^2(k)}{2} \ud k < 0, \\
  \nonumber
  e_{42}(g)\!\!&=&\!\!\!\!\int\limits_0^\infty\!\frac{F(k g)}{k}\frac{k^2 (J_1^2(k)\!-\!J_3^2(k))\!+\!8J_1(k)J_3(k)}{4 k^2}\ud k >0, \\
 \nonumber
  e_{44}(g)\!\!&=&\!\!-\!\!\int\limits_0^\infty\!\frac{F(k g)}{k}\frac{ J_1^2(k)+8J_2^2(k)-9 J_3^2(k)}{8}\ud k < 0,
\end{eqnarray} 
where $F(x)=1-[1-\exp(-x)]/x$ and $J_i(x)$ are the Bessel's functions of the first kind. They can be expressed in closed form through the complete elliptic integrals.

The exchange energy, calculated as $(C/2)\sum_i(\vec{\nabla} m_i)^2$, in the same normalization and order in $1/p$ is
\begin{equation}
 \label{eq:bendex}
 e_{EX}^1=\frac{1 + 2(p^2-2) x^2 + 4 x^4}{\gamma_B p^4 \rho^2},
\end{equation} 
where $\rho=R/L_E$.

The total energy $e^1=e_{EX}^1+e_{MS}^1$ is a polynomial in $1/p$ and $x$. It can be easily minimized analytically. The solution for equilibrium $p$ always exists and stable. For $x$, surprisingly, the equations gave two solutions. In certain geometries the solution $x=0$ (the ``leaf'' state) is stable, whereas in other geometries the energy minimum at $x=0$ turns into the maximum, and another solution at $|x|<1$ (which is a ``C''-type state, or a bent ``leaf'') becomes stable. For even larger cylinders the only stable solution has $x>1$, which is vortex. The critical lines ``e'' and ``f'' result from simultaneously solving three equations $\partial e/\partial p=0$, $\partial e/\partial x=0$ and $\partial^2 e/\partial x^2=0$. These equations are polynomial and the critical lines can be expressed in a closed form.

Consideration of vortex core expansion instability requires to allow for the core to be bigger than the particle size (the case not considered in Refs.~\citen{UP93}, \citen{UP94}). That is $p=1$, $a=0$ and $c<1$  (or $c=\cos\alpha$, $\alpha\in[0,\pi/2]$) in Eq.~(\ref{eq:fz}). Only the surface magnetic charges, $m_z(r)=2/(1+c^2 r^2)-1$, are present. Their normalized magnetostatic energy (the self-energy of both faces and their interaction) can be written as $e_{MS}^2=g_U(0,\alpha)-g_U(g,\alpha)$. The function $g_U$ is factored as
\begin{equation}
 g_U(h,\alpha)=\int\limits_0^\infty e^{- k h}\left[\int\limits_0^1 r
 \frac{1-r^2 \cos^2\alpha}{1+r^2 \cos^2\alpha}
 J_0(k r)\right]^2 \!\!\!\ud k.
\end{equation}The exchange energy after direct integration is
\begin{equation}
 \label{eq:eex2}
e_{EX}^2 = \frac{1}{\gamma_B\rho^2}\frac{8\cos^2\alpha}{3+\cos2\alpha}.
\end{equation} 
The critical line ``g'' was calculated numerically, by solving a system of equations: $\ud e^2/\ud \alpha =0$ and $\ud^2 e^2/\ud \alpha^2 =0$, where $e^2=e_{EX}^2+e_{MS}^2$.

This research was performed with financial support by KOSEF through the Quantum Photonic Science Research Center, Korea, and by MOST, Korea. Authors would like to thank Dr. Min Hyung Cho for reading the manuscript and many valuable discussions.

\end{document}